# Optical Calibration of a Frequency-Domain-Multiplexed X-Ray TES Microcalorimeter Array via Photon-Number-Resolved Statistics

**Giacomo Gorla [1,2], Bernard Sohl [3], Emanuele Taralli [4,*], Martin de Wit [4], Luciano Gottardi [4], Kevin Ravensberg [4], Kenichiro Nagayoshi [4], Davide Vaccaro [4], Hugo van Bohemen [3], Mio Poortvliet [3] and Wolfgang Löffler [3]**

1 INAF-IAPS, 00133 Rome, Italy; giacomo.gorla@inaf.it
2 Physics Department, Sapienza University of Rome, Piazzale Aldo Moro 2, 00185 Rome, Italy
3 Huygens-Kamerlingh Onnes Laboratory, Leiden Institute of Physics, Leiden University, 2311 EZ Leiden, The Netherlands; bohemen@physics.leidenuniv.nl (H.v.B.); poortvliet@physics.leidenuniv.nl (M.P.); loeffler@physics.leidenuniv.nl (W.L.)
4 NWO-I/SRON Space Research Organisation Netherlands, 2333 CA Leiden, The Netherlands; m.de.wit@sron.nl (M.d.W.); l.gottardi@sron.nl (L.G.); k.ravensberg@sron.nl (K.R.); d.vaccaro@sron.nl (D.V.)
* Correspondence: e.taralli@sron.nl

**Abstract**

Transition-edge sensor (TES) microcalorimeters for future X-ray space observatories require accurate low-energy calibration, a regime where conventional X-ray sources are limited by intrinsic linewidth, count rate, or experimental complexity. In this work, we investigate fiber-coupled optical photons as a compact calibration tool for an Athena-like TES array. Optical pulses at 405 nm were generated with an acousto-optic modulator and delivered through an optical fiber onto a selected TES pixel inside a dilution refrigerator, via a dedicated cryogenic alignment interface. Externally triggered pulses were processed with an optimal-filter pipeline, including baseline subtraction, drift correction, and multi-Gaussian reconstruction of the pulse-height spectrum. The reconstructed spectra show resolved photon-number peaks consistent with the Poissonian statistics expected from an attenuated coherent source, reaching equivalent energies above 400 eV with a mean photon number exceeding 100 per pulse. The detector response was further tested for linearity by comparing the measured photon-number distributions against those predicted from the independently calibrated number of photons sent to the detector and a single, globally estimated detection efficiency, confirming Poissonian, linear behaviour across nearly two orders of magnitude in photon number. These results demonstrate that fiber-delivered optical photons can probe the few-eV to sub-keV response of Athena-like TES microcalorimeters, offering a practical and extensible framework for future low-energy calibration studies.

Academic Editor: Ismael Martel Bravo

Received: 9 September 2026
Revised: 6 October 2026
Accepted: 13 October 2026
Published: 27 October 2026

## 1. Introduction

High-resolution X-ray spectroscopy is a key tool for investigating the hot and energetic Universe [1], including the dynamics [2], chemical enrichment [3], and thermodynamic state [4] of diffuse astrophysical plasmas. The X-ray Integral Field Unit (X-IFU) [5] on board NewAthena [6] is being developed to provide spatially resolved spectroscopy with less than 4 eV energy resolution over a broad soft-X-ray energy range. Achieving this performance requires large arrays of superconducting microcalorimeters with stable, reproducible, and accurately calibrated responses. Transition-edge sensors (TESs) are among the leading detector technologies for this purpose. A TES microcalorimeter [7] is operated within the narrow superconducting-to-normal transition, where a small temperature increase caused by photon absorption produces a measurable change in resistance. Under voltage bias, this resistance change is converted into a current pulse whose amplitude is related to the deposited energy. Negative electrothermal feedback stabilizes the detector operating point and improves the linearity and recovery of the response, enabling high-resolution spectroscopy. For TES arrays, multiplexed readout is essential to reduce the number of cryogenic wires and the associated thermal load. In frequency domain multiplexing (FDM) [8], each pixel is AC-biased through a dedicated superconducting LC resonator at a distinct carrier frequency. Photon-induced changes in each pixel's TES impedance modulate its carrier. The carriers from all pixels are combined into a single signal, amplified by one shared two-stage SQUID, and digitally demodulated back into the individual pixel channels. This FDM readout architecture enables the characterization and operation of individual pixels in large TES arrays while keeping power consumption low, since several tens of pixels can be multiplexed through the same amplification chain [9].

A remaining challenge is the calibration and characterization of TES microcalorimeters at low energies. Conventional fluorescent X-ray calibration lines below 1 keV can be broader than the intrinsic detector resolution [10,11]. Compact, lab-based monochromators are limited by the achievable photon flux, while synchrotron beamline monochromators avoid this limitation but require access to a large external facility. Electron-beam ion traps (EBITs) can likewise be coupled to TES microcalorimeters, as demonstrated for higher-energy transitions and dedicated atomic-physics studies [12,13], but similarly require large and dedicated infrastructure and provide low intensity at low energies. Besides, measurements below approximately 100–200 eV are challenging because the infrared-blocking filters, required in the cryostat to limit the thermal load on the detector, can significantly reduce the transmission of such low-energy X-rays. Another option is to use a multiwall carbon nanotubes to generate electrons with kinetic energy in the range of 100 eV [14], however this approach requires a substantial modification of the experimental setup.

Optical photons provide an attractive alternative [15,16]. When a coherent optical pulse is sufficiently attenuated, the number of absorbed photons follows Poisson statistics, generating a comb of peaks separated by the known optical photon energy, a technique recently used to cross-validate TES energy calibration against an independent X-ray source on a single detector [16,17]. In this work, we extend optical photon-number-resolved characterization to Athena-like Ti/Au TES microcalorimeters with full-size Au/Bi absorbers, operated with the SRON frequency-domain multiplexed (FDM) readout. Their larger heat capacity, compared with the smaller-absorber devices used in previous optical-photon demonstrations [15–17], reduces the pulse height per photon, making single- to multi-photon discrimination more challenging and more representative of the space mission. We demonstrate full energy scale Poissonian linearity up to approximately 400 eV and apply the same characterization procedure to three simultaneously read-out pixels, establishing compatibility with a multiplexed array configuration.

## 2. Materials and Methods

### *2.1. Experimental Setup*

#### 2.1.1. The TES Array

The measurements were performed on an Athena-like transition-edge sensor array fabricated at SRON, where TES pixels are coupled to large Au/Bi absorbers. The nominal pixel pitch is 317 μm, with absorber gaps of 5 μm and an absorber stack consisting of 1.0 μm Au and 5.5 μm Bi. The TES bilayer is based on Ti/Au, with a target sheet resistance of approximately 15 mΩ/sq [18]. In this work, we used high-ohmic 105 × 15 μm$^2$ TES geometry (Figure 1 on the left) developed to match the SRON FDM readout [19]. The 8 × 9 pixel array can be read out from one side of the chip, as shown schematically in the center of Figure 1, while a microscope image of the mounted detector assembly is shown on the right. The expected energy resolution for this array is 1.6 to 2.1 eV at 6 keV. The TES chip is wire-bonded to the cryogenic readout circuitry, including the micro-strip line interconnect chip and LC-filter stages, and is thermally anchored to the cold stage of the dilution refrigerator.

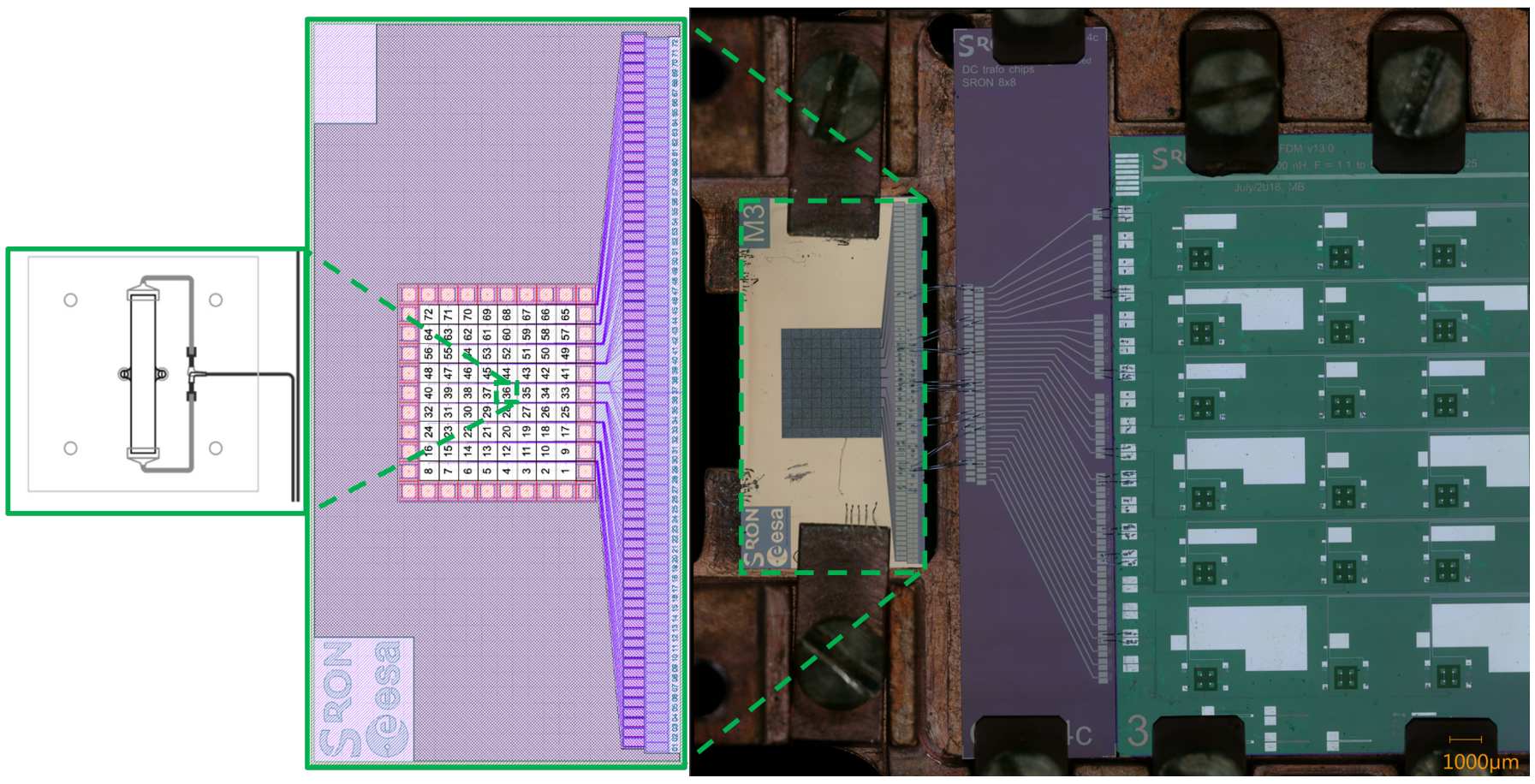

**Figure 1.** SRON measurement setup: (**Right**) Microscope image of the mounted TES detector assembly. The TES array is wire-bonded to the cryogenic readout chain, including the micro-strip line interconnect chip and LC-filter stages used for frequency domain multiplexed operation. (**Center**) Schematic layout of the 8 × 9 TES array, corresponding to 72 pixels, designed to be read out from one side of the chip. (**Left**) Single pixel geometry 105 µm× 15 µm.

For the optical calibration measurements, six pixels of the array were biased and read out simultaneously through the FDM system, each coupled to its own dedicated LC resonator and amplified by the SQUID readout chain. Pixel 36 was aligned with the optical fiber and was the intended illumination target. The detector was operated at a bath temperature of approximately 58 mK during the optical pulse measurements. The fiber was aligned to pixel 36 using a dedicated cryogenic mechanical interface, consisting of a custom aluminum superconducting shield and a copper alignment bracket mounted on the detector backbone (Figure 2 left). During assembly, the copper bracket held the cleaved fiber ferrule above the TES array while the shield remained open enough to inspect the relative fiber–absorber position optically. The ferrule's insertion depth into the copper block set its height above the TES chip, while the bracket allowed transverse (X and Y) adjustment until the fiber axis was centered on the target pixel's absorber through specific screws. Once aligned, the bare fiber was passed through a slit in the removable half of the three-part aluminum shield (Figure 2 right), which was then closed around the bracket and TES chip; only afterward was the chimney screwed onto the removable half to complete the feedthrough. This sequence avoids threading the fiber through the narrow chimney bore while it is already positioned, limiting mechanical stress on it during assembly. Once closed, the aluminum shield provides a light-tight enclosure that reduces stray illumination of pixels or internal components and becoming superconducting below about 1 K, it also provides magnetic shielding of the TES array during operation.

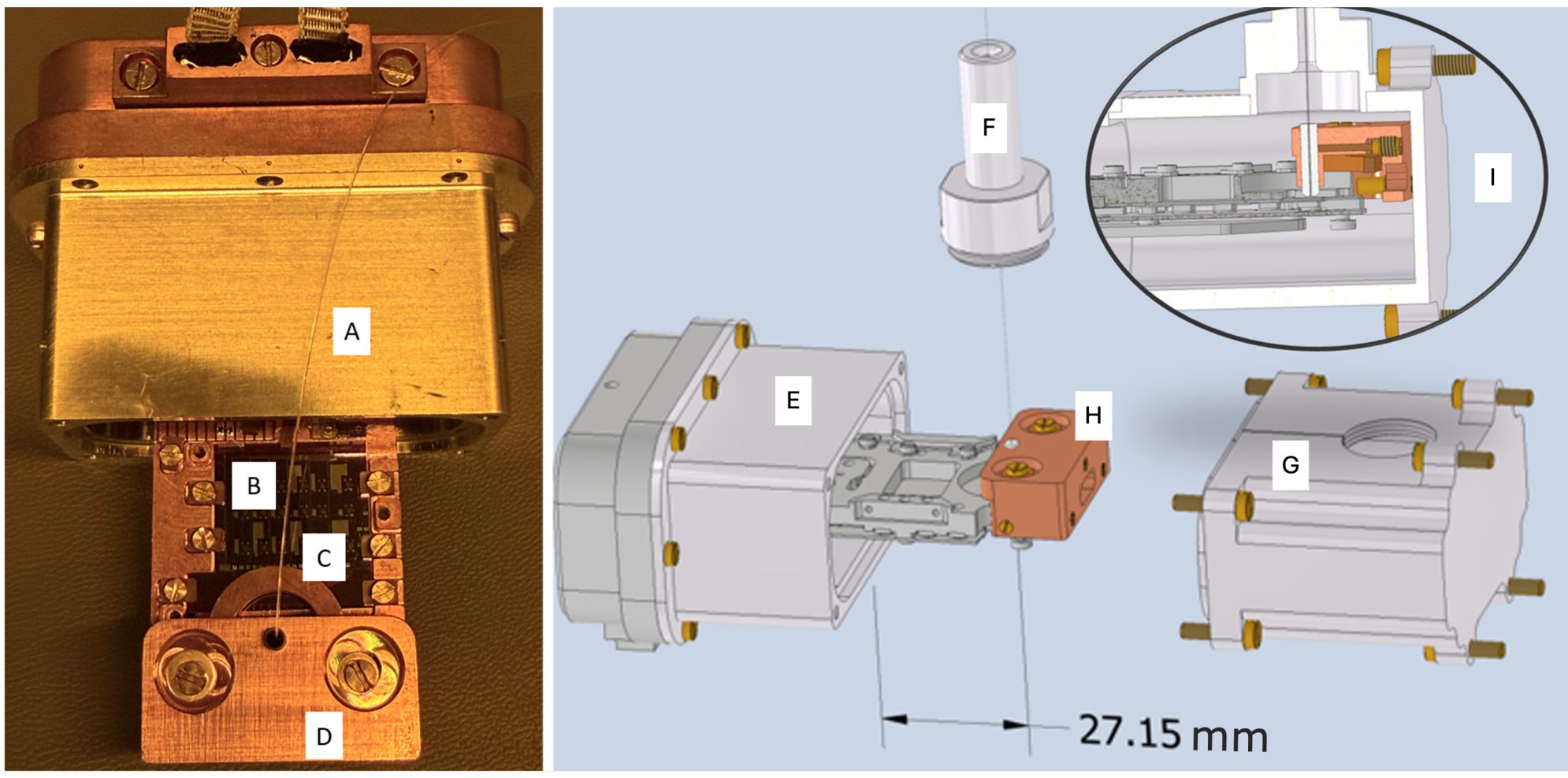

**Figure 2.** (**Left**) Cryogenic detector assembly with the optical fiber positioned toward the TES array. The dedicated mechanical interface allows the fiber to be aligned to a selected pixel before closing the assembly. A: Optical Fiber, B: LC Filter, C: Interconnector Chip, D: Copper Bracket. (**Right**) Exploded view of the aluminum three-part shield (E, F and G), copper detector backbone and fiber-alignment bracket (H). Inset I: Cross-sectional view of the assembled shield, showing the fiber ferrule routed through the chimney and held by the copper bracket close to the TES array.

### 2.1.2. Optical Setup

An acousto-optic modulator (AOM) generated light pulses with various averaged photon numbers. A schematic of the optical bench is shown in Figure 3. The light source was a 405 nm laser diode. At this wavelength, each photon carries an energy of approximately 3.06 eV, providing a natural few-eV calibration spacing when photon-number peaks are resolved in the TES pulse-height spectrum.

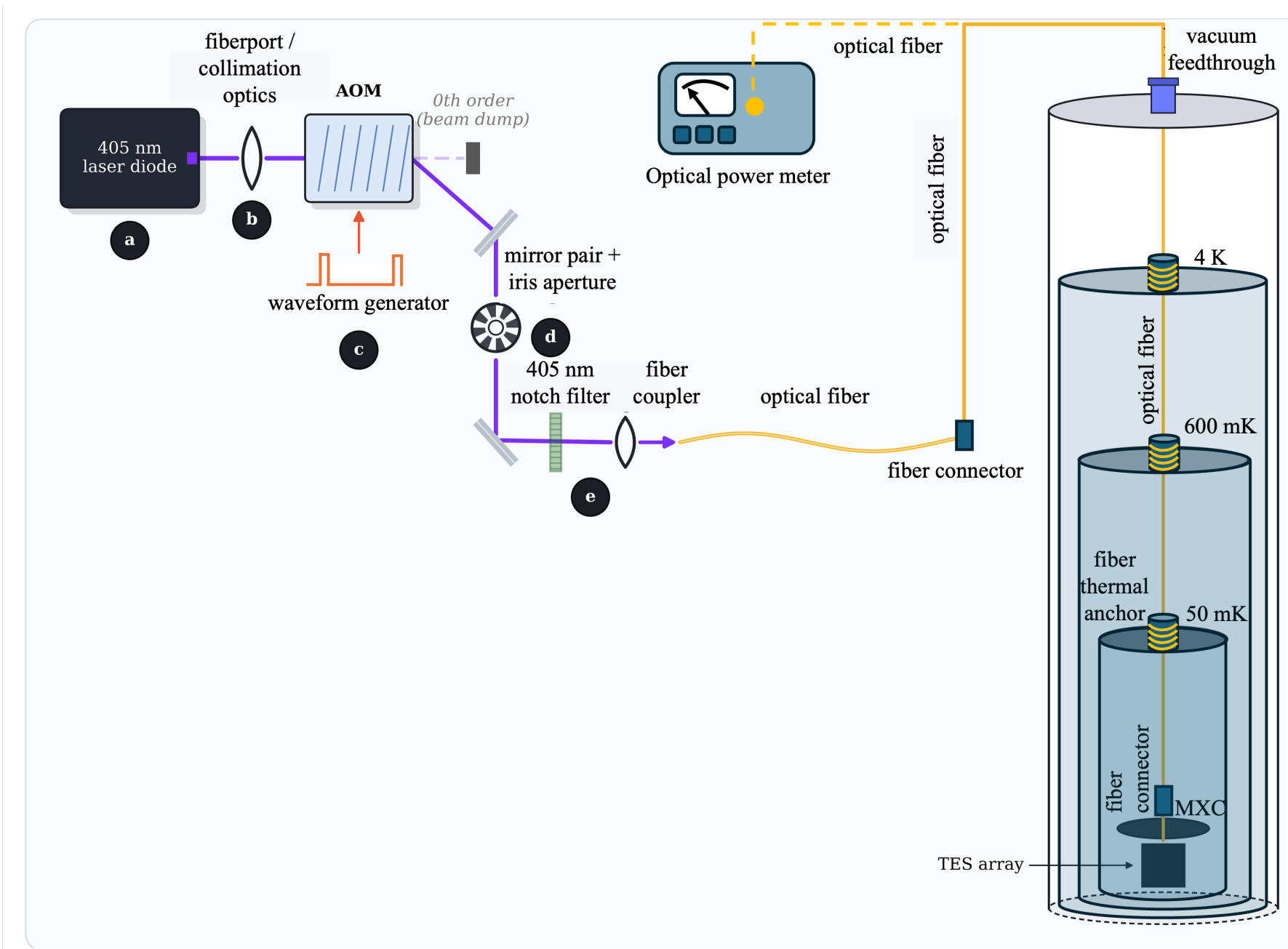

**Figure 3.** Optical pulsing setup used to generate fiber-coupled optical pulses. The main components are: (a) 405 nm laser diode, (b) fiberport/collimation optics, (c) acousto-optic modulator, (d) mirror pair and iris aperture used to isolate the first-order diffracted beam, and (e) 405 nm notch filter plus fiber coupler used to inject the pulsed light into the optical fiber.

The laser beam was first collimated and then sent through the AOM, driven by a waveform generator producing pulses with widths ranging from 200 ns to approximately 5 μs, repeated every 20 ms to fully comply with the detector dead time. Within a pulse-duration series, the laser-diode current was kept constant and the optical energy was varied by changing the AOM gate width. This duration range is considerably shorter than the characteristic TES response time, so the optical excitation can be regarded as effectively instantaneous with respect to the TES response. The drive signal excites acoustic waves in the piezoelectric crystal, creating a time-dependent diffraction grating that deflects the beam into a first-order component only while the drive is on. Selecting this component makes the AOM act as a fast optical shutter, with the same drive signal providing an external electrical trigger synchronized to each optical pulse. This scheme is essential for reaching the lowest photon numbers, since direct modulation of our laser driver cannot produce electrical pulses of sufficiently short duration. After the AOM, the diffracted beam was isolated using a mirror pair and an iris aperture, then coupled into an optical fiber and guided into the cryostat. The optical intensity was attenuated to reach the low photon-number regime required for photon-number-resolved measurements. Inside the cryostat, the fiber delivered the optical pulses to the TES array and was positioned onto the selected pixel using the cryogenic alignment interface described above. For the acquisitions requiring the highest mean photon numbers, the AOM stage was bypassed: the waveform generator directly modulated the laser diode current, and its output fiber was connected directly to a feedthrough on top of the cryostat.

### *2.2. Preliminary Analysis and Software Optimization*

The optical pulse datasets were analyzed using a custom Python3 pipeline developed for externally triggered TES measurements. For each acquisition, the software first averaged an initial subset of traces to identify the approximate pulse position and to define the pulse window. Samples outside this window were used to estimate the local baseline of each event. A robust baseline subtraction was applied trace by trace before amplitude reconstruction. The baseline was calculated from the pre- and post-pulse regions after rejecting the largest deviations from the local median, in order to reduce the effect of spurious pulses or glitches occurring outside the triggered window. This step ensures that the reconstructed pulse height is referred to a common zero level for all events. The pulse amplitude was reconstructed using an optimal filter in the frequency domain. In the analysis presented here, a fixed common pulse template was used for all datasets and is shown in Figure 4. This standard procedure ensures that the same amplitude estimator is consistently applied to all datasets. The template was generated from a representative high-signal-to-noise acquisition by selecting up to 1000 pulses belonging to a resolved photon-number population. The selected traces were baseline-subtracted, temporally aligned, and individually normalized in amplitude; pulses affected by pile-up, glitches, or anomalous baseline fluctuations were rejected. The final template was obtained from the median of the remaining traces and normalized to unit pulse height. It was then aligned vertically to the baseline and to the detected pulse position and Fourier-transformed. For each event, the Fourier transform of the baseline-subtracted trace was compared with the external template and weighted by the inverse noise power spectral density. The optimal-filter amplitude was computed as

$$A = \frac{\mathrm{Re}\left[\sum_f D(f)S^*(f)/J(f)\right]}{\sum_f |S(f)|^2/J(f)}, \tag{1}$$

where $D(f)$ is the Fourier transform of the measured trace, $S(f)$ is the Fourier transform of the external template, and $J(f)$ is the noise power spectral density. This procedure pro-vides a single pulse-height estimator for each triggered event while suppressing frequency components dominated by noise.

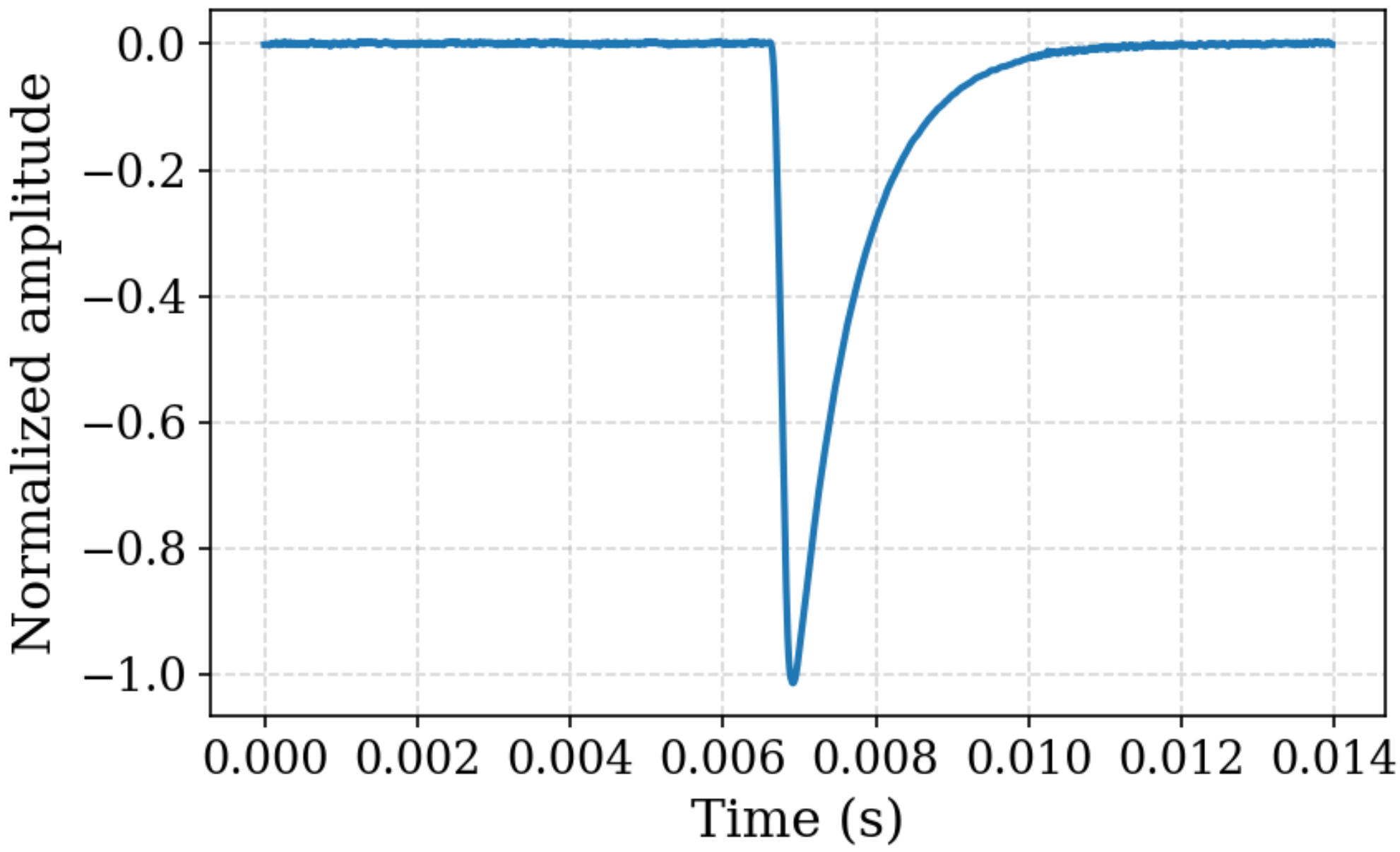

**Figure 4.** Normalized template obtained via the optimal filtering of the custom pipeline.

As a robustness check, we repeated the analysis using a template generated independently for each dataset. The resulting peak positions, energy calibration, and energy-resolution values were consistent with those obtained using the common template within the corresponding statistical uncertainties. After optimal filtering, a baseline-drift correction was applied to compensate for slow variations in detector response during acquisitions. The correction was obtained by evaluating the correlation between the local pre-pulse baseline and the reconstructed pulse amplitude. A linear trend was fitted to the densest region of this distribution, thereby avoiding outliers, and the corresponding correction was applied to all events. The corrected pulse amplitudes were then histogrammed to reconstruct the photon-number spectrum. The spectra were fitted with a multi-Gaussian model in pulse-height units. The zero-photon component was treated independently, while the photon peaks were constrained to follow a common spacing. The fitted peak areas were used to extract the photon-number probabilities and compare them with the Poisson distribution expected for an attenuated coherent optical source. Finally, the fitted spacing between adjacent photon-number peaks was used to calibrate the pulse-height axis in energy, using the known photon energy of the 405 nm laser.

## 3. Results

Optical pulse acquisitions were performed for different laser settings to probe the pixel response over a broad range of absorbed optical energies. The reconstructed spectra show a clear comb-like structure, with adjacent peaks separated by the energy of a single 405 nm photon. Representative energy-calibrated spectra for a subset of the optical pulse acquisitions analyzed in this work are shown in Figure 5 (left panels), together with the corresponding photon-number probability distribution (right panels).

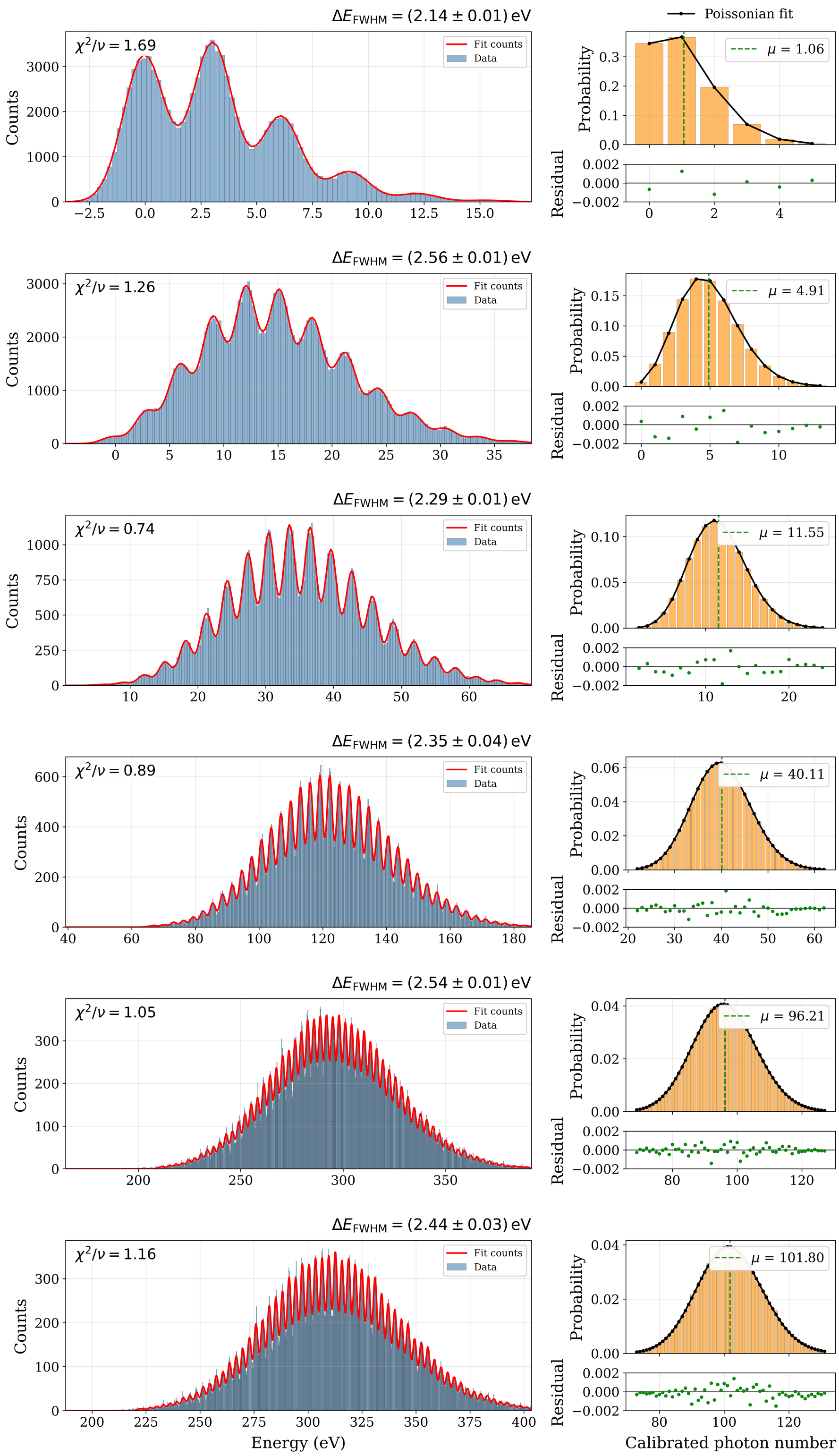

**Figure 5.** Representative energy-calibrated pulse-height spectra for a subset of the optical pulse settings analyzed in this work, spanning fitted mean photon numbers from $\mu_{\text{meas}} = 1.06$ to $\mu_{\text{meas}} = 101.80$. For each dataset shown, the measured spectrum (histogram) is shown together with the multi-Gaussian fit and the evaluated $\Delta E_{\text{FWHM}}$ with its propagated $1\sigma$ uncertainty (**left panels**), while the extracted photon-number probabilities are compared with the Poisson distribution expected for an attenuated coherent source (**right panels**). Residuals between the measured probabilities and the Poisson fit are reported alongside each panel.

For low energy optical pulses, only the first few photon-number peaks are significantly populated, while increasing the laser current or pulse duration shifts the Poisson distribution toward larger photon numbers. The fitted mean photon number ranges from approximately $\mu = 1.06$ to $\mu = 101.80$ in the datasets shown. In all cases, the extracted photon-number probabilities follow the expected Poissonian statistics of an attenuated coherent optical source, with residuals from the Poisson fit remaining $< 0.002$ in absolute probability over the full range, confirming that the observed peak structure reflects photon-number statistics rather than readout artifacts. Both cross-talk between channels and background events would appear as excess counts or off-spacing peaks inconsistent with Poisson statistics, neither of which is observed. This is consistent with prior measurements on arrays from the same fabrication process, which showed a first-neighbour thermal crosstalk below $10^{-4}$ for a 6 keV pulse [20,21], and, for the background, scaling the rate measured on an array with smaller absorbers [22] to our absorber area gives an upper limit of $\lesssim 5 \times 10^{-7}$ counts $s^{-1}$ over 0–400 eV, i.e., less than one event in about 23 days.

The multi-Gaussian fits reproduce the measured spectra over a wide dynamic range, with a fitted energy resolution of approximately 2.1 to 2.5 eV FWHM for the datasets in Figure 5, which is in line with the expected energy resolution of the TES array. For each dataset, $\Delta E_{\mathrm{FWHM}}$ is obtained from the common Gaussian width of the resolved photon peaks; the first fitted component has a separate width and is excluded from the reported resolution. The width in pulse-height units, $\sigma_{\mathrm{PH}}$, is converted to energy using the fitted peak spacing $\Delta_{\mathrm{PH}}$, according to $\Delta E_{\mathrm{FWHM}} = 2.355\,\sigma_{\mathrm{PH}}\,(3.06\,\mathrm{eV}/\Delta_{\mathrm{PH}})$, and the reported $1\sigma$ uncertainty is propagated from the covariance matrix of the multi-Gaussian fit and the energy calibration. The quality of the individual fits was assessed using the reduced $\chi^2_\nu = \chi^2/\nu$, computed with the actual number of free parameters in the constrained model. For the spectra in Figure 5, $\chi^2_\nu$ ranges from 0.74 to 1.69, confirming that the common-width, equally-spaced multi-Gaussian description accurately reproduces the data within the adopted Poisson counting uncertainties. The reported FWHM is related to the corresponding Gaussian energy standard deviation by $\Delta E_{\mathrm{FWHM}} = 2\sqrt{2\ln 2}\,\sigma_E \simeq 2.355\,\sigma_E$. The degradation in energy resolution at higher photon numbers is visible from the $\Delta E_{\mathrm{FWHM}}$ values reported in the panel labels and is consistent with the filling-in of valleys between adjacent peaks observed in other experiments [15,23], possibly caused by extra noise from laser light incident on parts of the chip near the aligned pixel. This shows that the TES pixel resolves individual photon-number states from the single-photon regime up to significantly higher absorbed energies, with peaks observed up to approximately 400 eV, corresponding to more than 130 absorbed 405 nm photons. This upper bound reflects a practical limitation of the current optical setup, rather than a fundamental limit of the detector itself.

The linearity of the detector response was evaluated directly in terms of photon-number statistics, rather than through the energy-scale fit alone, using the mean number of photons independently sent to the detector for each acquisition, $\mu_{\mathrm{sent}}$. This quantity was obtained from power-meter measurements of the laser output, taken at pulse widths and diode currents matching each acquisition, with the power meter positioned immediately before the fiber entered the cryostat (see Figure 3), so the derived detection efficiency describes the full transmission from the cryostat input to the TES pixel. Two Poisson references are compared against each measured distribution. The first, $q(n) = \mathrm{Poisson}(n; \mu_{\mathrm{meas}})$ (yellow bars in Figure 6a), uses each dataset's own fitted mean and checks only whether the measured distribution has the expected Poissonian shape, a self-Poisson comparison, simpler and less stringent than the full POVM (Positive Operator-Valued Measure) reconstruction used by Brida et al. [24]. The second, $l(n) = \mathrm{Poisson}(n; \hat{\eta}\,\mu_{\mathrm{sent}})$ (blue bars in Figure 6a), uses a single, globally estimated detection efficiency, $\hat{\eta} = \sum_j \mu_{\mathrm{meas},j} / \sum_j \mu_{\mathrm{sent},j} = 2.02 \times 10^{-5}$, and is the direct test of detector linearity: because $\hat{\eta}$ is fixed across all datasets, agreement with $l(n)$ shows that one efficiency predicts the detected mean across nearly two orders of magnitude in sent photon number. The low value of $\hat{\eta}$ mainly reflects the intentional splice losses inside the cryostat, but upstream of the aluminum shield, from the 200 µm core delivery fiber down to a 5 µm and finally a 3 µm core fiber illuminating the TES. Reflective losses at the Bi absorber surface at 405 nm also concur to the low efficiency and may contribute to the degradation in energy resolution at high photon number.

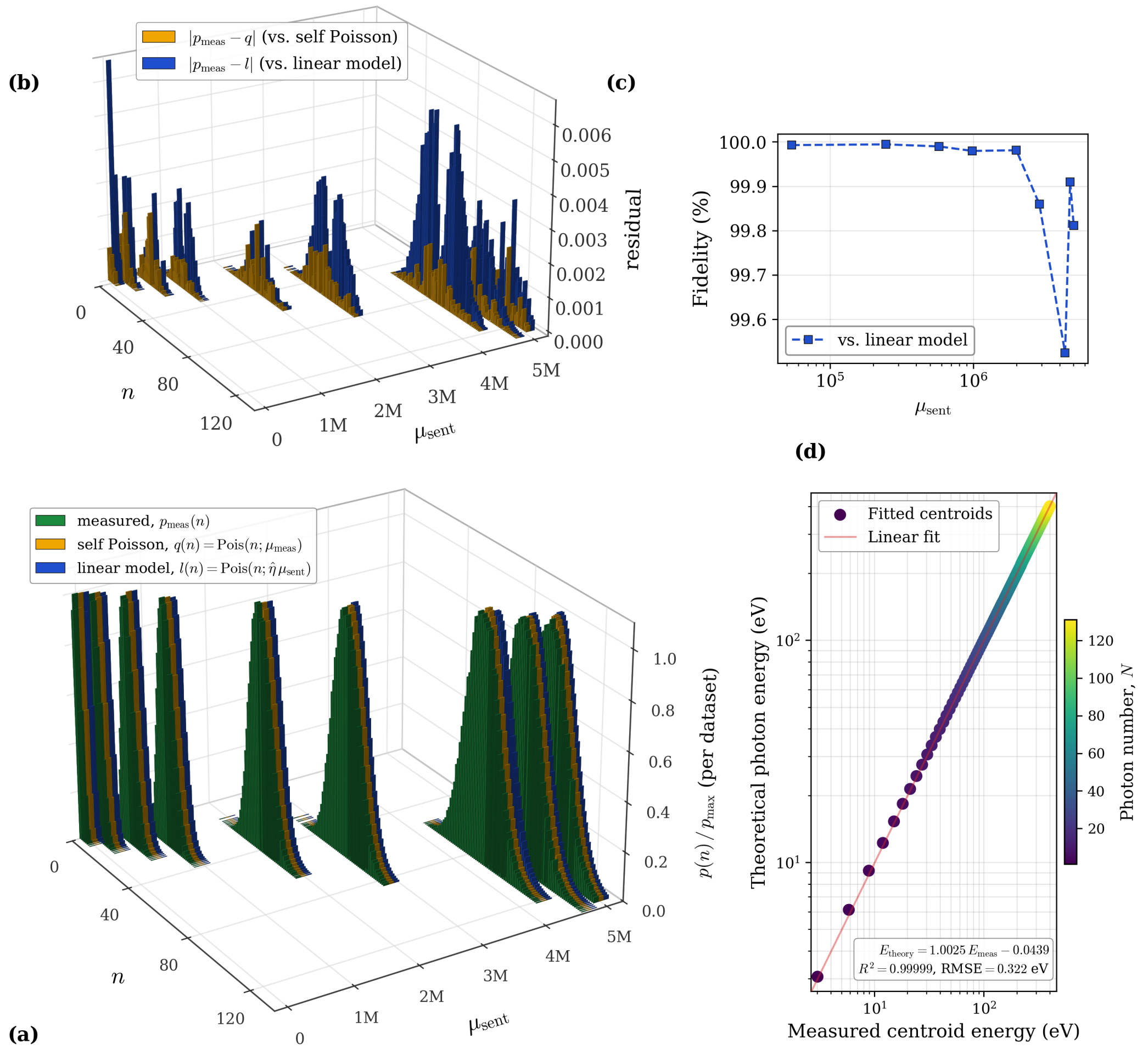

**Figure 6.** Linearity comparison across the optical pulse datasets. (**a**) Measured (green), self-Poisson (gold), and linear-model (blue) photon-number distributions. (**b**) Absolute residuals between the measured distribution and each model. (**c**) Fidelity between the measured and linear-model distributions vs. mean sent photon number. (**d**) Fitted centroid energies vs. theoretical photon energies, an independent linearity check. See main text for details.

Despite this low overall efficiency, the measured photon-number distributions remain in good agreement with the Poisson distribution expected for an attenuated coherent source across the optical pulse datasets, both when compared to each dataset's own fitted mean and to the distribution predicted from $\mu_{sent}$ and the detection efficiency, consistent with the TES response remaining Poissonian and linear from $\mu \approx 1$ up to $\mu \approx 102$. Figure 6b shows the absolute residuals, $|p_{meas}(n) - q(n)|$ and $|p_{meas}(n) - l(n)|$, between the measured distribution and each of the two models. As a result of using a single, fixed efficiency across all datasets, the residuals from the linear model are systematically larger than those from the self-Poisson comparison. Figure 6c shows the fidelity between the measured distribution and the linear model, $F = \sum_n \sqrt{p_{meas}(n)\, l(n)}$, as a function of the mean sent photon number. Figure 6d gives a complementary and more direct linearity check: the fitted

Gaussian centroid energies (points, color-coded by photon number $N$) are compared with the expected photon energies, $E_N = NE_\gamma$, where $E_\gamma$ is the 405 nm photon energy, together with a linear fit (red line). When a given photon number appeared in more than one dataset, the centroid with the highest statistics was retained. Unlike panels (a) to (c), which test the linearity of the full photon-number probability distribution against an independently known sent flux, panel (d) tests only whether the fitted peak positions themselves scale linearly with photon number, independent of any power-meter calibration. The measured centroids follow the theoretical photon-energy scale with excellent linearity over the full investigated range ($E_{\text{theory}} = 1.0025\,E_{\text{meas}} - 0.0439$, $R^2 = 0.99999$, RMSE = 0.322 eV).

Figure 7 summarizes the nine optical pulse settings included in the stacked energy-calibrated histogram. Each coloured contribution corresponds to one acquisition condition, making visible how the different pulse widths and laser powers populate complementary photon-number ranges. The stacked representation is used here as a compact overview of the full calibration data set: the same photon-energy spacing is preserved across the acquisitions, while the logarithmic scale shows that the optical-pulse method remains usable from the single-photon regime to the highest absorbed energies investigated.

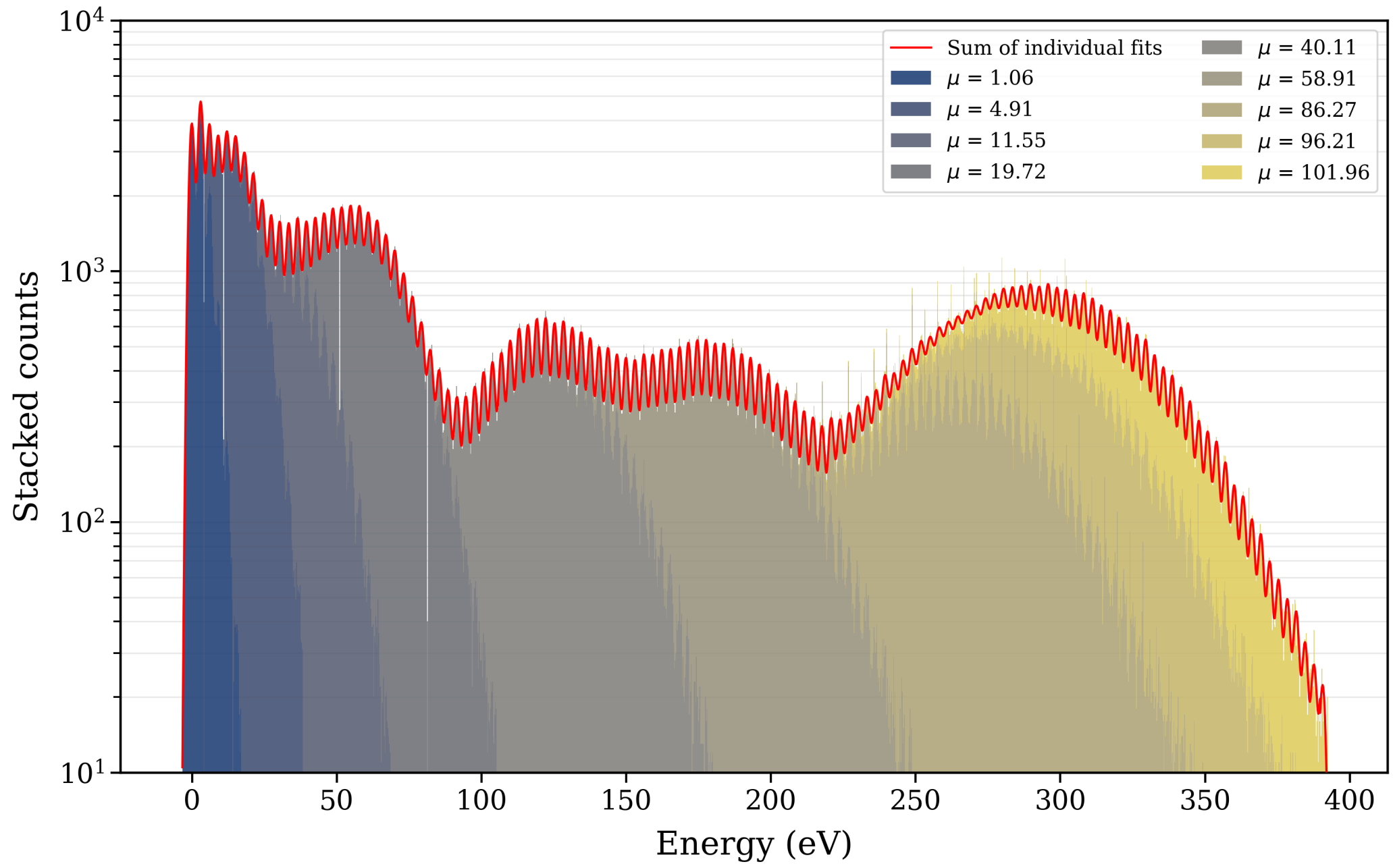

**Figure 7.** Stacked energy-calibrated histograms for the nine optical pulse settings included in this overview. Each colour corresponds to one dataset, showing how the different acquisition conditions cover complementary photon-number and energy ranges on a logarithmic count scale. The sum of the individual fit (red line) serves only as a guide to the eye.

*Multiplexed Optical Response from Multiple Pixels*

Two additional TES pixels detected optical photons, although the fiber had been aligned to a single pixel. The spectra of the three channels, labelled Px36, Px44, and Px35, were each reconstructed and energy-calibrated with their own fitted peak spacing and optimal-filter template. As shown in Figure 8a, all three exhibit resolved photon-number peaks and photon-number distributions described by Poisson fits, despite receiving different mean photon numbers, with no localized excess or second comb that would indicate a fixed-amplitude pickup artifact or crosstalk between channels. A stronger, independent test is whether the three separate calibrations agree with each other. We combined the three spectra into a single histogram (Figure 8b) and fit it with one global multi-Gaussian model. The resulting resolution, $\Delta E_{\text{FWHM}} = (2.61 \pm 0.27)$eV, matches that of the individual datasets (Figure 5): any relative miscalibration or template bias between pixels would have broadened the combined peaks well beyond this level, yet the comb remains fully resolved after summing. This confirms that the method extends to simultaneous, multiplexed readout of several pixels without measurable cross-pixel calibration bias.

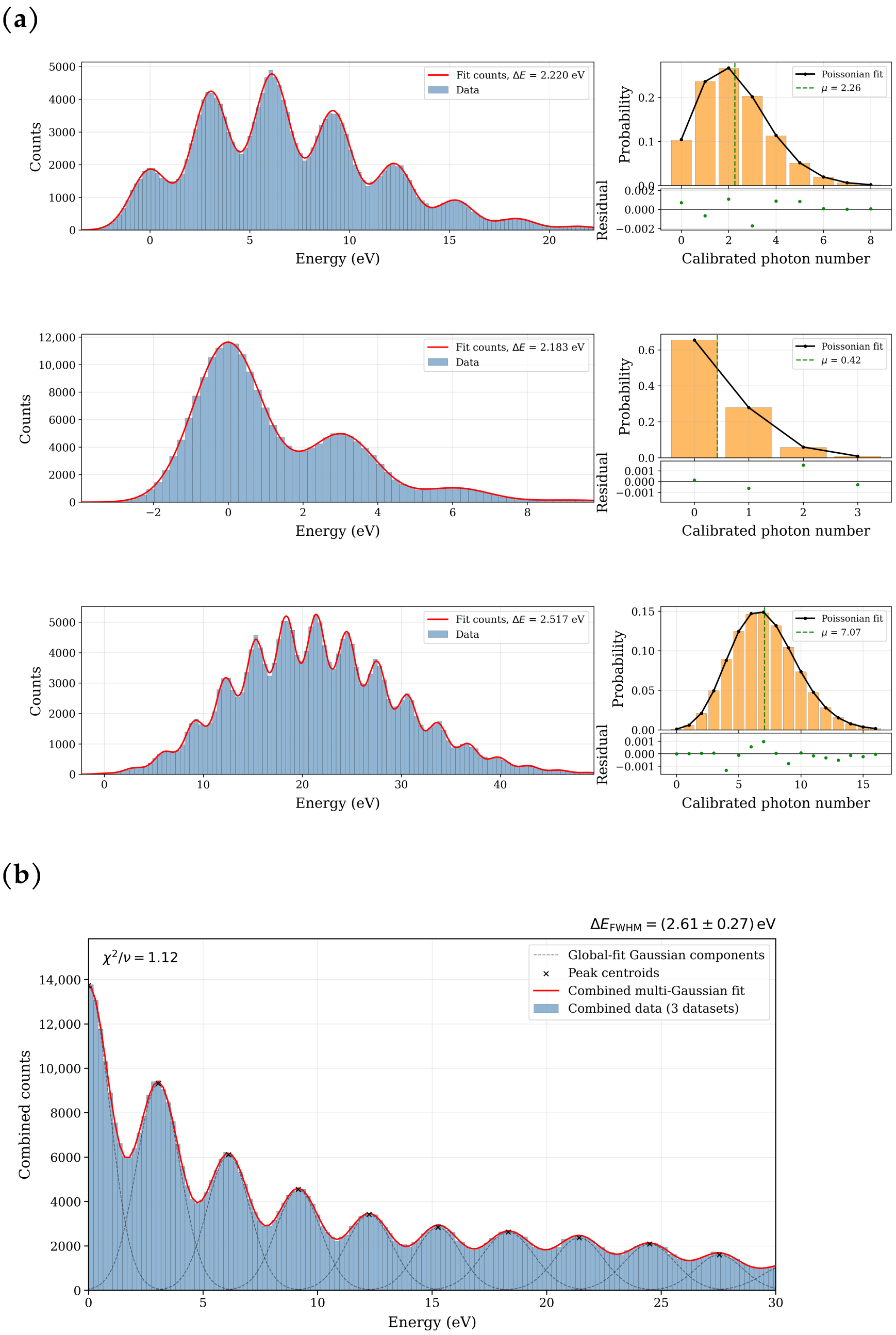

**Figure 8.** Simultaneous optical response of three FDM-multiplexed TES pixels. (**a**) Individual energy-calibrated spectra and fits (**left**), and photon-number distributions with Poisson fits and residuals (**right**). The rows correspond, from top to bottom, to Px36, Px44, and Px35, respectively. The reported $\Delta E_{\mathrm{FWHM}}$ values include the corresponding propagated $1\sigma$ uncertainties. (**b**) Combined energy-calibrated histogram obtained from all three pixels, fit with a single global multi-Gaussian model in which the centroid and width of each component are fitted independently.

## 4. Conclusions

This work demonstrated fiber-delivered 405 nm optical photons as a compact calibration and characterization method for an Athena-like TES microcalorimeter array fabricated at SRON and operated with the SRON FDM readout. Externally triggered pulses were processed with a custom optimal-filter pipeline based on an external pulse template, robust baseline subtraction, baseline-drift correction, and multi-Gaussian reconstruction of the photon-number spectra. Across the optical pulse datasets, the reconstructed photon-number distributions remain in good agreement with Poisson statistics, both against each dataset's own fitted mean and against the mean sent photon number scaled by a single global detection efficiency. This confirms that the detector response remains Poissonian and linear from the single-photon regime up to $\mu \approx 101.80$, with the fitted peak spacing matching the 405 nm photon energy throughout and an energy resolution in the few-eV range. These results establish fiber-coupled optical photons as a flexible tool for the low-energy characterization of Athena-like TES arrays. The observation and quantitative characterization of three simultaneously read-out pixels further demonstrate the compatibility of the method with multiplexed array operation. Future work will focus on improving the optical confinement onto selected absorbers and comparing the optical calibration with independent X-ray calibration measurements.

**Author Contributions:** Conceptualization, B.S., E.T. and H.v.B.; Methodology, E.T.; Software, G.G. and B.S. and K.N.; Validation, G.G., E.T., L.G., M.P. and W.L.; Formal analysis, G.G., B.S. and E.T.; Investigation, B.S., E.T., M.d.W. and L.G.; Resources, K.R., D.V., H.v.B. and M.P.; Data curation, G.G., B.S. and E.T.; Writing—original draft, G.G. and E.T.; Writing—review & editing, G.G., E.T., M.d.W., L.G., K.R., D.V., H.v.B., M.P. and W.L.; Visualization, G.G., B.S. and E.T.; Supervision, E.T. and W.L.; Funding acquisition, L.G. and W.L. All authors have read and agreed to the published version of the manuscript.

**Funding:** This research received no external funding.

**Institutional Review Board Statement:** Not applicable.

**Informed Consent Statement:** Not applicable.

**Data Availability Statement:** Data are available from the corresponding author upon reasonable request.

**Acknowledgments:** The authors acknowledge the support of the SRON cryogenic detector and readout teams and the Leiden University optics laboratory for their contributions to the optical setup and detector testing activities. Bernard Sohl, Hugo van Bohemen, Mio Poortvliet and Wolfgang Löffler acknowledge funding from NWO (680.92.18.04), NWO/OCW (Quantum Software Consortium Nos. 024.003.037 and 024.003.037/3368, and from the Dutch Ministry of Economic Affairs (QDNL).

**Conflicts of Interest:** The authors declare no conflicts of interest.

## Abbreviations

The following abbreviations are used in this manuscript:

| | |
|---|---|
| AOM | Acousto-optic modulator |
| FDM | Frequency domain multiplexing |
| FWHM | Full width at half maximum |
| SQUID | Superconducting quantum interference device |
| TES | Transition-edge sensor |
| X-IFU | X-ray Integral Field Unit |